\documentclass{emulateapj}

\newcommand{\cmpss}{cm~s$^{-2}$}
\newcommand{\mps}{m~s$^{-1}$}
\newcommand{\mpstwo}{m~s$^{-2}$}
\newcommand{\kps}{km~s$^{-1}$}

\newcommand{\Msun}{${\rm M_\odot}$}
\newcommand{\Rsun}{${\rm R_\odot}$}
\newcommand{\Mjup}{${\rm M_J}$}
\newcommand{\xonb}{XO-2b}
\newcommand{\xon}{XO-2}
\newcommand{\Rjup}{${\rm R_J}$}
\newcommand{\vMs}{0.98}		
\newcommand{\eMs}{0.02}
\newcommand{\vRs}{0.97}		
\newcommand{\epRs}{0.02}
\newcommand{\enRs}{0.01}
\newcommand{\sptype}{K0V}
\newcommand{\vrvK}{85} 
\newcommand{\ervK}{8}

\newcommand{\vDs}{150}		
\newcommand{\epDs}{4}
\newcommand{\enDs}{2}

\newcommand{\vjd}{2454147.74902}	
\newcommand{\ejd}{0.0002}	
\newcommand{\vap}{0.0369}	
\newcommand{\eap}{0.0002}

\newcommand{\vperiod}{2.615857}	
\newcommand{\eperiod}{0.000005}
\newcommand{\vMp}{0.57}
\newcommand{\eMp}{0.06}	
\newcommand{\vRp}{0.98}		
\newcommand{\epRp}{0.03}
\newcommand{\enRp}{0.01}
\newcommand{\vincl}{88.9}
\newcommand{\eincl}{0.7}

\newcommand{\vAge}{5.3}
\newcommand{\epAge}{1.0}
\newcommand{\enAge}{0.7}
\newcommand{\vSepas}{31}
\newcommand{\vFeH}{0.45}
\newcommand{\eFeH}{0.02}
\newcommand{\vgp}{14.8}
\newcommand{\egp}{1.6}

\slugcomment{Accepted for publication in the Astrophysical Journal}

\received{2007 Sep 14}
\begin{document}

\title{XO-2\lowercase{b}: Transiting Hot Jupiter in a Metal-rich Common Proper Motion Binary}

\author{
Christopher~J.~Burke\altaffilmark{1}
P.~R.~McCullough\altaffilmark{1},
Jeff~A.~Valenti\altaffilmark{1},
Christopher~M.~Johns-Krull\altaffilmark{2},
Kenneth~A.~Janes\altaffilmark{3},
J.~N.~Heasley\altaffilmark{4},
F.~J.~Summers\altaffilmark{1},
J.~E.~Stys\altaffilmark{1},
R.~Bissinger\altaffilmark{5},
Michael~L.~Fleenor\altaffilmark{6},
Cindy~N.~Foote\altaffilmark{7},
Enrique~Garc\'{i}a-Melendo\altaffilmark{8},
Bruce~L.~Gary\altaffilmark{9},
P.~J.~Howell\altaffilmark{3},
F.~Mallia\altaffilmark{10},
G.~Masi\altaffilmark{11},
B.~Taylor\altaffilmark{3},
T.~Vanmunster\altaffilmark{12}
}

\email{cjburke@stsci.edu}

\altaffiltext{1}{Space Telescope Science Institute, 3700 San Martin Dr., Baltimore MD 21218}
\altaffiltext{2}{Dept. of Physics and Astronomy, Rice University, 6100 Main Street, MS-108, Houston, TX 77005}
\altaffiltext{3}{Boston University, Astronomy Dept., 725 Commonwealth Ave.,Boston, MA 02215}
\altaffiltext{4}{Inst. for Astronomy, University of Hawaii, 2680 Woodlawn Dr., Honolulu, HI 96822-1839}
\altaffiltext{5}{Racoon Run Observatory, Pleasanton, CA}
\altaffiltext{6}{Volunteer Observatory, Knoxville, TN}
\altaffiltext{7}{Vermillion Cliffs Observatory, Kanab, UT}
\altaffiltext{8}{Esteve Duran Observatory Foundation, Montseny 46, 08553 Seva, Spain}
\altaffiltext{9}{Hereford Arizona Observatory, Hereford, AZ}
\altaffiltext{10}{Campo Catino Astronomical Observatory, Guarcino, Italy}
\altaffiltext{11}{Virtual Telescope Project, Bellatrix Astronomical Observatory, Ceccano, Italy}
\altaffiltext{12}{CBA Belgium Observatory, Landen, Belgium}

\begin{abstract}
We report on a V=11.2 early K dwarf, \xon\ (GSC 03413-00005), that hosts a
$R_{p}$=\vRp$\pm^{\epRp}_{\enRp}$ \Rjup, $M_{p}$=\vMp$\pm$ \eMp\ \Mjup\
transiting extrasolar planet, \xonb, with an orbital period of
\vperiod$\pm$\eperiod\ days.  \xon\ has high metallicity,
[Fe/H]=\vFeH$\pm$\eFeH, high proper motion, $\mu_{tot}=157$
mas~yr$^{-1}$, and has a common proper motion stellar companion with
\vSepas $\arcsec$ separation.  The two stars are nearly identical
twins, with very similar spectra and apparent magnitudes.  Due to the
high metallicity, these early K dwarf stars have a mass and radius
close to solar, $M_{\star}=\vMs\pm
\eMs$ \Msun\ and $R_{\star}=\vRs\pm^{\epRs}_{\enRs}$ \Rsun .  The high
proper motion of \xon\ results from an eccentric orbit (Galactic
pericenter, $R_{per}<4$ kpc) well confined to the Galactic disk
($Z_{max}\sim 100$ pc).  In addition, the phase space position of \xon\
is near the Hercules dynamical stream, which points to an origin of
\xon\ in the metal-rich, inner Thin Disk and subsequent dynamical
scattering into the solar neighborhood.  We describe an efficient
Markov Chain Monte Carlo algorithm for calculating the Bayesian
posterior probability of the system parameters from a transit light
curve.
\end{abstract}


\section{Introduction}

We announce the discovery of an extrasolar planet, \xonb, that
transits a bright, V=11.2, star.  With an orbital period, $P\sim 2.6$
days, planetary radius, $R_{\rm p}=$\vRp\ \Rjup, and planetary mass,
$M_{\rm p}=$\vMp\ \Mjup,
\xonb\ belongs to the growing class of transiting Hot Jupiter (HJ)
planets \citep{CHA07}.  Despite the increasing number of transiting
planets known, much work still remains to understand the observed
properties of planets.  The transiting planets HD209458b
\citep{CHA00,HEN00}, HAT-P-1b \citep{BAK07}, and WASP-1b \citep{CAM07,CHA07b} have anomalously large
radii compared to theoretical models and are thought to require an
external source of energy to remain inflated
\citep{BOD03,GUI02,WIN05}.  However, recently \citet{BUR06} explain
the radii of transiting planets without the need for an extra source
of energy by accounting for enhanced metallicity opacities and
properly comparing observed radii to theoretical radii.  In contrast,
HD149026b has an extremely high density and small radius
\citep{SAT05}.  There is general agreement that a pure H/He mixture
cannot explain the small radius of HD149026b, but a planet model
with a massive central core, $M_{c}\sim 70\ M_{\oplus}$, of heavy
elements along with a small H/He envelope can explain the
radius of HD149026b
\citep{SAT05,FORT06}.

Disentangling the effects of stellar irradiation, migration, central core mass,
and composition on the observed properties of planets requires discovering
more bright transiting planets \citep{CHA07,FORT06}.  \xonb\ is the
second contribution to the bright transiting planet sample provided by
the XO Project \citep{MCC05}; XO-1b being the first \citep{MCC06}.  Of
the bright transiting planet hosts ($V<12$), \xon\ has the highest metallicity ([Fe/H]=\vFeH\ see
\S~\ref{sec:sme}).  In addition, from a recent catalog of extrasolar planets
\citep{BUT06}, \xon\ has a higher metallicity than 96\% of all known
extrasolar planet hosts.

Metallicity plays a crucial
role in the formation of planets and the resulting HJ atmospheres.
The frequency of radial velocity detected planets is known to increase
with metallicity \citep{FIS05}.  In the core accretion planet
formation model, a high metallicity environment grows larger cores
and enables more objects to reach the critical mass necessary for
runaway gas accretion and transformation into a detectable gas giant
planet \citep{IDA04,BENZ06}.  Detailed fits to the mass and radius of
the known transiting planets yields a mass estimate for the central
refractory element core \citep{GUI06,BUR06}.  These investigations
derive a larger core mass for planets that transit higher
metallicity stars.  In addition to the bulk properties of extrasolar
planets, metallicity plays an important role in the planet's
atmosphere, especially for a HJ experiencing large stellar irradiation.
An increased metallicity results in a greater absorption of the stellar
irradiation and larger equilibrium temperatures than a comparable
planet of lower metallicity \citep{FORT06}.  The variations in equilibrium
temperature can lead to variations in the dominant observable features
in the planet's atmosphere.

Amongst the other transiting HJ planets, \xonb\ shares a common
characteristic with another transiting HJ, HAT-P-1b \citep{BAK07}.
Both planets orbit one member of a nearly-equal-mass, wide separation
stellar binary.  The \xon\ stellar binary system has a separation of
\vSepas $\arcsec$ ($\sim$ 4600 AU separation with a distance of $\sim$
150 pc) and both components have an identical within the uncertainties
157 mas\ yr$^{-1}$ proper motion vector as measured with Tycho-2
\citep{HOG00} and the high proper motion catalog of
\citep{LEP05}.  It is not unusual for planets to exist in binary
systems.  \citet{RAG06} find $>$23\% of stars with planets have a
stellar companion, however, they find evidence that the extrasolar
planet sample is deficient in stellar binaries when compared to the
field.  A stellar companion with a 4600 AU separation is not expected
to influence the planet formation process.  Even assuming an orbital
eccentricity, $e$=0.8, for the unknown eccentricity of the
\xon\ stellar binary, planets within $a_{c}\sim 170$ AU of \xon\ are
dynamically stable \citep{HOL99} (we assume the current measured
projected binary separation is close to the actual semi-major axis of
the orbit when calculating $a_{c}$).  Empirically, \citet{DES07} do not
find any statistically significant difference between planets around
single stars and planets in wide ($a>100$ AU) separation binaries.
However, understanding the influence of a stellar companion on the
planet formation process is complicated by the fact that the binary
configuration during planet formation may be vastly different than
what is currently observed \citep{MAR07,MAL07}.

The similar brightness of the \xon\ stellar binary components and
their angular separation provides an excellent opportunity for
detailed line abundance studies.  The higher metallicity of the known
planet host stars is thought to be of primordial origin rather than
heavy element pollution due to infalling planets \citep{FIS05,GON06}.
The infalling planet model for explaining the higher metallicity of
the known planet host stars has predominately come from observations
showing differing element abundances between common proper motion
binary components.  These element abundance differences have typically
been overturned by further independent analyses
\citep{GON06}.  The difficulty in reliable abundance differences
results from systematic effects in analyzing stars of different
$T_{eff}$ and evolutionary state \citep{SCH06}.  Such effects are
reduced for the \xon\ stellar binary.

\begin{figure}
\plotone{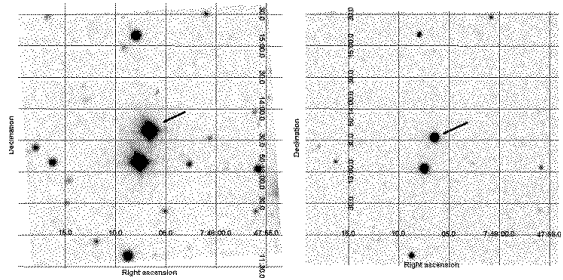}
\caption{Finder chart for \xon\ and its stellar companion at two epochs.  The arrow indicates \xon.  North is toward the top of the figure and East is toward the left of the figure.  ({\it Left}) Digital Sky Survey POSS I Red image
from 1953.  ({\it Right}) 2MASS H-band image from 2000.
The common proper motion is predominately directed South.\label{finder}}
\end{figure}

On the sky, \xon\ and a line joining its stellar companion has a
position angle of 162$\degr$, nearly along the
North-South direction (see the finder chart in Figure~\ref{finder}).
\xon\ is the Northern component of the binary (indicated by the arrow
in Figure~\ref{finder}).  The magnitude difference between the
components is $\Delta V=0.05$ mag.  Thus, in the optical, \xon\ is
fainter than its Southern companion.  The traditional nomenclature for
binary stars designates \xon\ as \xon B and the planet
\xon Bb.  However, throughout this article we simply designate the
transiting HJ stellar host \xon\ and the planet around it
\xon b.  When it is unclear from the context we alternatively designate
\xon\ as \xon N and the Southern stellar companion as \xon S.  We adopt
this naming convention because the magnitude difference between the
stellar components is small and the spectra are nearly identical, both
of which make it difficult to distinguish the objects, whereas the
\vSepas $\arcsec$ separation readily distinguishes the stellar
components based on declination.  We refer to both stellar components
as a unit as the \xon\ stellar binary.

In \S~\ref{sec:obs}, we provide details of the discovery and follow up
photometry and spectroscopy.  The high resolution spectroscopy is
analyzed in \S~\ref{sec:sme} to determine the stellar properties of
\xon N and \xon S.  The high precision photometry is analyzed in
\S~\ref{sec:lcmcmc} employing an efficient Markov Chain Monte Carlo (MCMC) algorithm to
determine the properties of \xonb.  We confirm the planetary mass of
\xonb\ with radial velocity measurements in \S~\ref{sec:rv}.  The
ephemeris for \xonb\ is refined and the transit observations are
investigated for transit timing variations in \S~\ref{sec:ttv}.  There
is a brief discussion regarding the \xon\ stellar binary and its
Galactic orbit in \S~\ref{sec:disc}, and we summarize the article in
\S~\ref{sec:sum}.

\section{Observations}\label{sec:obs}

\subsection{XO Project Photometry}

\xonb\ is the second transiting planet discovered by the XO survey after
XO-1b \citep{MCC06}.  \citet{MCC05} and \citet{MCC07} describe the
instrumentation, operation, and analysis in more detail than the
summary provided here.  The twin, 200mm XO cameras power drift scan in
declination over a $7\deg \times 62\deg$ strip every ten minutes on
clear nights for more than 2 months per season of visibility.  The XO
observations employ a broad (0.4$\mu$m to 0.7$\mu$m) passband.  The
star \xon\ comes from the XO strip centered on RA 8.0 hr and is one of
several thousands of bright (V$<12$) stars monitored in this strip
over the period September 2003 to September 2005.  The Box-fitting
Least Squares algorithm, BLS, \citep{KOV02} was employed to search the
nearly 3000 observations per star for repetitive transit events with
periods ranging from $P=0.5-10$ day.  We perform the transit search on
two realizations of the light curve.  One realization is the
calibrated light curves as described in \citet{MCC05} and the other
realization of the light curve has the SysRem algorithm \citep{TAM05}
applied to the calibrated light curve in order to further remove
systematics.  In the case of \xon, there were no substantive
improvements in the light curve quality after implementing SysRem and
both analyses identified the same transit period and phase.

\citet{MCC07} describe the selection of transit candidates for
followup.  In addition, for consideration the candidates must pass
selection criteria as described in \citet{BURK06}.  In brief, we
require observations covering more than 1.5 transit events, avoid 0.5
and 1.0 day periods where systematic aliases result in false-positive
detections, require the transit depth, $\delta m<0.1$ mag, and require
the transit to have higher significance than systematic errors (as
measured by the transit to anti-transit Ratio Statistic, RS, of
\citet{BURK06}).

\begin{figure}
\plotone{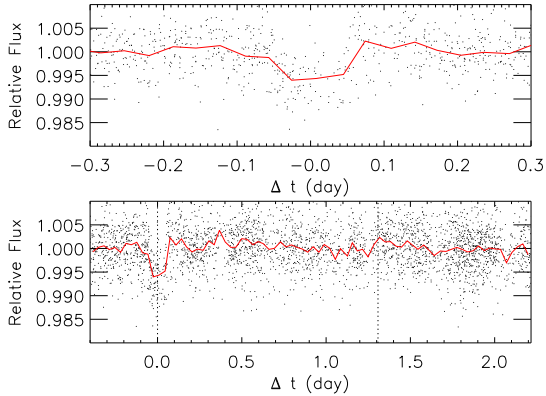}
\caption{Discovery phased light curve from the XO Project data.    ({\it Top}) The phased light curve around the transit event.  ({\it Bottom}) The phased light curve over the full orbital period.  The individual measurements ({\it points}) are shown binned ({\it solid line}) to reduce noise.  The transit occurs at $\Delta t=0.0$ day and any secondary eclipse for a circular orbit would occur at $\Delta t=1.3$ day ({\it dotted line}).  No secondary eclipse is evident above the noise.  At the resolution of the XO cameras, both components of the \xon\ stellar binary are within the photometry aperture causing dilution of the transit depth.\label{xophased}}
\end{figure}

Figure~\ref{xophased} shows the XO light curve phased on the detected
period as returned from BLS.  We achieved 0.7\% or 0.007 mag per
observation precision on this V=11.2 object.  The transit occurs at
$\Delta t=0.0$ day and any signature of a secondary eclipse would
occur at $\Delta t=1.3$ day for a circular orbit.  At the angular
resolution of the XO survey, both components of the \xon\ stellar
binary are within the photometry aperture leading to transit depth
dilution.  Light curves resolving the components of the \xon\ stellar
binary result in twice the transit depth (see Figure~\ref{et436}), but
in the discovery light curve the transit depth is 0.007 mag.  With the
XO cameras, we observed two nearly complete transits of \xonb\ at the end
of 2004, and beginning of 2005 (Julian dates 2453355 and
2453376).  In addition, on seven other nights, XO cameras captured
partial ingress and egress events.  First occurrence of a partial
transit was at the end of 2003 on Julian date 24552994.
Table~\ref{table:lc} provides a sample of the photometry for \xon\
from the XO cameras.  The full table is available in the online
edition.

\subsection{Extended Team and Follow Up Photometry}\label{sec:etphot}

\begin{figure}
\plotone{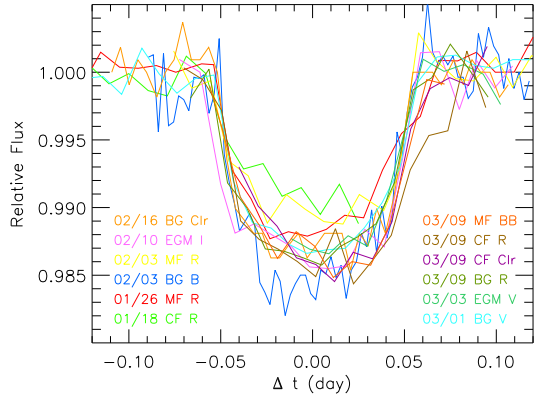}
\caption{Binned light curves from the Extended Team for \xon.  The color of the text in the lower left corner indicates the date (2007), observer, and passband of the observations.  The passband labels are Johnson for the system except {\it Clr} indicates unfiltered observations and {\it BB} indicates a blue blocking ($>$ 0.5 $\micron$) filter. \label{et436}}
\end{figure}

The Extended Team (E.T.) provides photometric follow up for XO
candidates.  The E.T.\ (R.~B., M.~F., C.~F., E.~G-M.,B.~G., P.~H.,
F.~M., G.~M., and T.~V.) is a collaboration of professional and
amateur astronomers \citep{MCC05,MCC06}.  We sent the candidate list
containing \xon\ to the E.T.\ on January 16, 2007.  On, January 18,
2007, E.T.\ observations confirmed the transit events for \xonb\ in the
R-band using a 0.6-m telescope (solid green line in
Figure~\ref{et436}).  The initial observations were truncated due to
high air mass, but the observations confirmed the XO ephemeris, the
transit occurs in the Northern star of the \xon\ stellar binary, and
the undiluted transit depth is consistent with a planetary radius and
the XO photometry.  A complete transit was observed by the E.T. on
January 26, 2007 and the binned R-band light curve obtained with a
0.35-m telescope is shown as the solid red line in Figure~\ref{et436}.
The light curves shown in Figure~\ref{et436} are used for refining the
ephemeris and looking for evidence of transit timing variations in
\S~\ref{sec:ttv}.  Based on observations of other transit candidates
by the E.T., repeat transit events in the same passband have a 0.2\%
or 0.002 mag standard deviation in deriving the transit depth.  At
this level of precision, the E.T.\ observations of
\xonb\ are consistent with a gray transit.  Table~\ref{table:lc}
provides E.T. photometry for \xon.  For the E.T. light
curves, the median differential magnitude out of transit provides the
flux normalization and the standard deviation out of transit provides
the uncertainty in the measurements.

\begin{figure}
\plotone{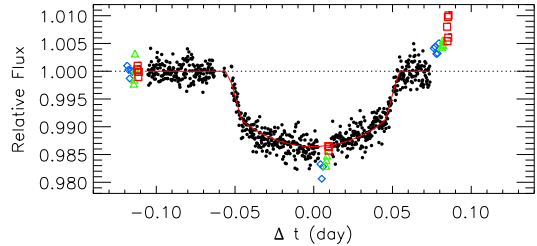}
\caption{Light curve from the 1.8m Perkins Telescope at Lowell Observatory in the R-band ({\it points}) for \xon N using \xon S as the comparison star.  The open symbols show the differential photometry in the B ({\it diamond}), V ({\it triangle}), and I ({\it square}) passbands using the average of the BVI observations obtained before the transit for the flux normalization level ({\it dotted line}).   Accurate positioning of the stellar image on the detector was not maintained during the BVI observations resulting in 0.5\% or 0.005 mag systematic offsets.  However, accurate positioning of the stellar image on the detector was maintained during the R-band light curve resulting in $\sim$0.0014 mag rms over 1 minute intervals.  Also shown is the best-fit transit model in a $\chi^{2}$ sense during the MCMC analysis. ({\it solid line}). \label{janeslc}}
\end{figure}

On February 16, 2007 we observed a transit event of \xonb\ with the
1.83-m Perkins Telescope at Lowell Observatory using the PRISM
instrument in imaging mode \citep{JAN04}.  The PRISM camera is a
2048x2048 Fairchild CCD with 0.39$\arcsec$/pix resolution.  The
transit event was well positioned in the evening, occurring over the
airmass range 1.05$\lesssim X\lesssim$1.16.  To improve efficiency a
subframe containing \xon N and \xon S was read out with a 8-10 s
cadence.  The average seeing during the course of observations,
3$\arcsec$, was poor for the site.  The final light curve is a
differential light curve for \xon N using \xon S as the comparison
shown in Figure~\ref{janeslc} employing aperture photometry and an
R-band filter.  The selected aperture size for photometry minimizes
the resulting rms scatter in the differential light curve.  In
addition to the R-band photometry, photometry in the BVI passbands was
obtained before the transit, at mid-transit, and after the transit.
The gap in the R-band data at mid-transit accommodates the multiple
filter data.  A computer failure resulted in the gap of R-band data at
the start of ingress.

To normalize the differential light curve, the average magnitude from
out of transit data was subtracted from the light curve.  This was
done independently for each half of the light curve separated by the
mid-transit gap.  There is a $\Delta m=0.001$ magnitude difference in
the normalization zeropoints between the data before and after the
transit.  This offset results from repositioning the telescope and
refocusing for the BVI data acquisition at mid-transit.  During an
uninterrupted R-band data series, a 0.3 pix rms positional accuracy was
maintained resulting in $\sim$0.0014 mag rms over 1 minute intervals.
Accurate positioning was not maintained for the BVI data resulting in
$\sim \Delta 0.005$ mag offsets between the data obtained before and
after the transit.  The average BVI passband data before transit was
subtracted from the BVI data obtained at mid-transit and after transit
to yield the differential light curve in these passbands as shown in
Figure~\ref{janeslc}.  Within the systematics resulting from
repositioning of the stars on the detector, the transit is gray.

We obtained photometric B, V, ${\rm R_C}$, and ${\rm I_C}$ magnitudes
for \xon N and \xon S using a 0.35-m telescope on
the photometric night of January 24, 2007 (Table~\ref{table:star}).  A single
Landolt area \citep{LAN92} was observed at the same airmass as \xon\ to
derive the zero point and color transformation coefficients.  The
color transformation coefficients were consistent with comprehensive
standard star measurements from four Landolt fields obtained two weeks
previous using the identical instrumental setup.  The color range of
Landolt standards was $-0.14\leq {\rm (B-V)} \leq 1.4$.  The B, V,
${\rm R_C}$, and ${\rm I_C}$ absolute photometric accuracies are 0.04,
0.04, 0.04, and 0.05 mag r.m.s., including both the rms scatter around
the photometric transformation model and an estimated systematic
error.  The Tycho-2 magnitudes for \xon\ listed in Table~\ref{table:star} transform
(via Table 2 of \citet{BES00}) to Johnson $V = 11.25$, i.e. 0.07 mag
(2-$\sigma$) fainter than our estimate.
In addition, we accurately measured the instrumental magnitude difference between \xon N and \xon S.  We find \xon N is fainter than \xon S by 0.07$\pm$0.008 mag in the B-band, 0.055$\pm$0.004 mag in the V-band, 0.040$\pm$0.004 mag in the R-band, and 0.030$\pm$0.003 mag in the I-band.

\subsection{Spectroscopy}\label{sec:spectroscopy}
After confirmation of the XO transit light curve from E.T.\
observations (see \S\ref{sec:etphot}), we initiated queue schedule
observations of \xon N and \xon S with the High-Resolution
Spectrograph (HRS), a fiber fed cross-dispersed echelle spectrograph
\citep{TUL98}, on the McDonald Observatory 11-m Hobby-Eberly Telescope
(HET) in order to measure the mass of the planet.  The first HRS
observations using an iodine gas cell for precision radial velocities
commenced on January 26, 2007.  Table~\ref{table:rv} provides dates of
the HRS observations along with the resulting radial velocities.  The
instrument setup provides R=60,000 resolution and wavelength coverage
over the range $4000<\lambda<7800$ $\AA$ with center at $\lambda=5900$
$\AA$.  We extracted the two-dimensional echelle spectra using
procedures described in \citet{HIN00}.  The resulting Signal-to-Noise
ratio (SNR) varied from 20-50 per extracted pixel at the blaze peak.
We calculate radial velocities for \xon N and \xon S in
\S~\ref{sec:rv}.

To measure the stellar parameters of \xon N and \xon S, we also
obtained spectra with the 2dCoud\'{e} echelle spectrometer
\citep{TUL95} on the McDonald Observatory 2.7-m Harlan J.\ Smith
Telescope (HJS).  We obtained two spectra of \xon N and a single
spectrum of \xon S with R=60,000 and wavelength coverage of 3900-9600
$\AA$.  We determine the stellar parameters in \S~\ref{sec:sme}.

\section{Analysis}

\subsection{Stellar Properties}\label{sec:sme}

We use the Spectroscopy Made Easy (SME) analysis package of
\citet{VAL96} with refinements from \citet{VAL05} on the HJS spectra
to measure the stellar properties of \xon N and \xon S.  We briefly
describe the process here.  The free parameters, $T_{eff}$, log$g$,
[M/H], $v\sin{i}$, [Na/H], [Si/H], [Ti/H], [Fe/H], and [Ni/H], are
varied in order to minimize the difference of the resulting synthetic
spectrum to the observed spectrum.  A quadratic continuum is fit over
8 wavelength intervals for each unique set of the above free
parameters.  When generating the synthetic spectrum, the
pressure-temperature profile of the atmosphere comes from
interpolating the atmosphere grid of \citet{KUR92}.  The atomic line
list comes from the Vienna Atomic Line Database (VALD) \citep{PIS95}
and the molecular line list comes from \citet{KUR93}.  The line
strengths and van der Waals damping parameter of the line list
database were adjusted to improve agreement with the observed solar
spectrum, as described in \citet{VAL05}.

\begin{figure}
\plotone{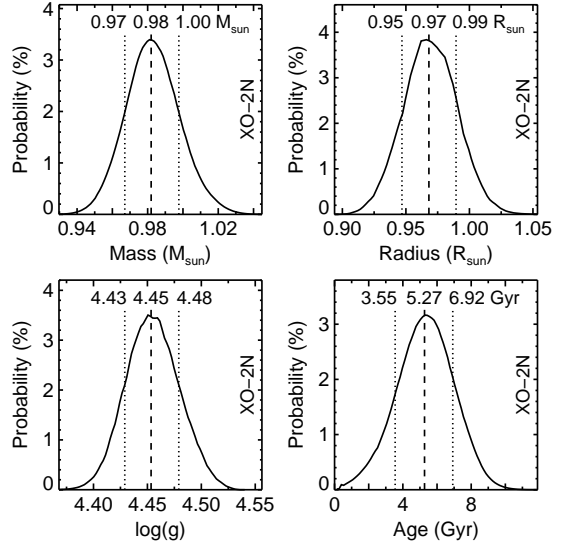}
\caption{Distributions for four stellar parameters derived from the
SME analysis (see text) for a distance of 150 pc. The values of
the mean and limits containing $\pm34$\%\ of the distribution from the mean
are annotated on the figures and listed in
Table~\ref{table:sme} along with
corresponding values for distances of 140 pc and 170 pc.\label{fig:smeiso}}
\end{figure}

Table~\ref{table:sme} lists the stellar parameters for XO-2N and XO-2S
based on the SME analysis.  The results for \xon N are based on the
average from two spectra and the results for \xon S are based on a
single spectrum.  Both \xon N and \xon S independently result in a
metal enhanced abundance, [Fe/H]=\vFeH$\pm$\eFeH.  Using the primary
observables from the SME analysis ($T_{eff}$, abundances, and log$g$)
and the apparent magnitude in the V-band (\S~\ref{sec:etphot}), we
determine secondary stellar properties, $M_{\star}$, $R_{\star}$,
distance, and age using the Y$^{2}$ isochrones \citep{YI01} following the procedure
of \citet{VAL05}.  The distance to \xon\ is unknown, thus the
probability density function for $M_{\star}$, $R_{\star}$, and age are
calculated for a sequence of trial distances in steps of 10 pc.  The SME
isochrone analysis for select distances to \xon\ are listed in
Table~\ref{table:smeiso}.  We determine a distance, d=\vDs\ pc, to
\xon\ in \S~\ref{sec:lcmcmc} from a joint analysis of the SME analysis
and transit light curve.  We show the probability density for the \xon
N parameters in Figure~\ref{fig:smeiso} for the preferred distance to
\xon\ of \vDs\ pc.

The spectrum alone yields an estimate of the stellar gravity,
log$g_{sme}$.  Estimates of $M_{\star}$ and $R_{\star}$ from the
isochrone analysis provide an additional estimate of log$g_{iso}$ as a
function of distance to \xon.  The condition log$g_{sme}$=log$g_{iso}$
yields an approximate distance to \xon .  The data for \xon S provides
a consistency check, and ideally the condition
log$g_{sme}$=log$g_{iso}$ for \xon S is met for the same distance as
\xon N.  In practice, we find \xon S has log$g_{sme}=4.6$; too large a
value for log$g_{iso}$ to accommodate at any distance.  Also, \xon S
being brighter and thus more massive than \xon N implies log$g_{sme}$ for
\xon S should be lower than log$g_{sme}=4.5$ of \xon N, opposite of what is
measured.  Analysis of the transit light curve along with the physical
parameters for \xon N from the isochrone analysis in
\S~\ref{sec:lcmcmc}, yields log$g_{lc,iso}\sim 4.5$, confirming the
spectroscopic log$g_{sme}$ for \xon N is correct and the log$g_{sme}$
for \xon S is an overestimate.

The overestimate in log$g_{sme}$ for \xon S is not without precedent.
The SME based log$g$ can result in higher values than the Y$^{2}$
isochrones allow for some objects in the Spectroscopic Properties of
Cool Stars (SPOCS) catalog as a result of the numerous degeneracies
that exist between the spectral parameters (see Figure 16 of
\citet{VAL05}).  Despite individual cases of an overestimated log$g$,
comparing stars analyzed with SME in common with other independent
spectral analyses does not reveal a systematic offset in log$g$ (see Figure 20
of \citet{VAL05}).  The 1-$\sigma$ uncertainties in the stellar
parameters given in Table~\ref{table:sme} are based on the typical rms
scatter in parameters measured in independent, multiple spectra for
stars in the SPOCS catalog.  Based on the 1-$\sigma$ uncertainties in
Table~\ref{table:sme}, the difference in log$g$ between \xon N and
\xon S is significant.  However, the distribution of repeat
measurements for parameters in the SPOCS catalog (see Table 5 of
\citet{VAL05}) display extended wings and the Gaussian-based
uncertainties underestimate the possibility of outlying measurements.
Thus, we also provide the 99.7\% confidence intervals for the
parameters based on Table 5 of \citet{VAL05}.  Since, the \xon S
parameters are based on a single spectrum, the 99.7\% confidence
interval is more appropriate than relying solely on the 1-$\sigma$
error when deciding on the significance of any differences between the
properties of \xon N and \xon S.  

\subsection{Markov Chain Monte Carlo Light Curve Analysis}\label{sec:lcmcmc}

\begin{figure}
\plotone{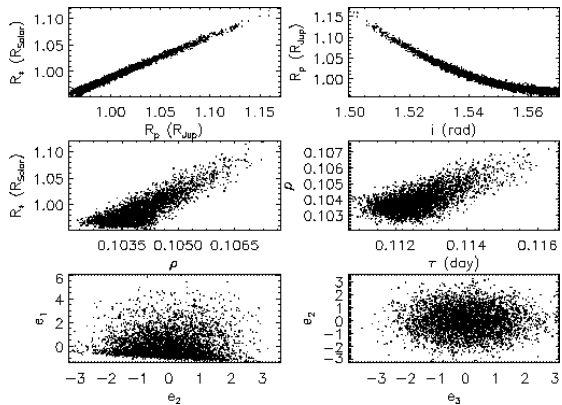}
\caption{Correlation between parameters for samples in a MCMC calculation of the \xonb\ posterior distribution.  The parameters $R_{\star}$, $R_{p}$, and $i$ are tightly correlated and the $R_{p}$ versus $i$ correlation is nonlinear ({\it Top}).  With the choice of parameters, $R_{\star}$, $\rho$, and $\tau$, the correlations between parameters are linear ({\it Middle}).  Solving for the eigenvectors of the covariance matrix from samples shown in the Middle Panel yields an eigenbasis set of parameters ({\it Lower}).  The eigenvector along the direction of largest variance, $e_{1}$ is predominately along $R_{\star}$.  Eigenvectors $e_{2}$ and $e_{3}$ are a mixture of $\rho$ and $\tau$.  The multi-normal without covariance properties of the eigenbasis set of parameters provides a more efficient MCMC calculation.\label{tranfig}}
\end{figure}

\citet{FORD05}, \citet{GRE05}, and references therein provide a
thorough discussion of the theory behind Markov Chain Monte Carlo
(MCMC) Bayesian analysis along with a practical MCMC implementation
for radial velocity planet detection.  In a multidimensional problem,
the MCMC algorithm is an efficient means of calculating the Bayesian
posterior probability for parameters.  \citet{HOL06} describe MCMC
analysis applied to determining the system properties for transiting
extrasolar planets.  In their analysis of the transit light curve for
XO-1b, \citet{HOL06} calculate the posterior probability using the
stellar radius, $R_{\star}$, planet radius, $R_{p}$, orbital
inclination, $i$, and the two coefficients of the quadratic limb
darkening law as free parameters.  In the transit fitting problem, the
relationship between $R_{\star}$, $R_{p}$, and $i$ has a high degree
of degeneracy and nonlinearity.  We illustrate the degeneracy and
nonlinearity between parameters in the top panels of
Figure~\ref{tranfig} for a MCMC calculation of the \xonb\ system
properties.  The nonlinear, ''banana-shaped'' degeneracy between
$R_{p}$ and $i$ slows the rate of convergence for MCMC algorithms.
MCMC algorithms work more efficiently when the relationship between
parameters is multi-normal without covariance
\citep{KOS02}.

To remove the nonlinearity between parameters, we choose the following
set of free parameters: $R_{\star}$, $\rho=R_{p}/R_{\star}$, and the
total transit duration from 1$^{\rm st}$ to 4$^{\rm th}$ contact,
$\tau$.  With this nonlinear transformation, the degeneracy between
the new set of parameters is along a straight line as shown in the
middle panels of Figure~\ref{tranfig}.  A further linear
transformation between parameters yields an eigenbasis set of
parameters with a multi-normal non-covariant relationship that results
in an efficient MCMC calculation.  These transformations improve the
rate of convergence for the MCMC analysis by more than a factor of 100
when measured by the autocorrelation length of the samples in the MCMC
sequence.  Appendix~\ref{apx:mcmc} provides further details of the
MCMC implementation developed for this study.

We employ the Metropolis-Hastings algorithm with a normal proposal
distribution for calculating the Markov Chain.  We follow a Gibbs-like
sampling technique where each step in the chain consists of a number
of intra steps updating each individual parameter in turn.  Several
short, trial chains iteratively yield scale factors of the normal
proposal distribution for each parameter with a 25\% to 40\%
acceptance rate for the trial samples.  The likelihood function is
given by $e^{-0.5\chi^{2}}$, were we have assumed the errors are
normally distributed, and the data have uniform weights.  $\chi^{2}$
is the squared difference between observations and the analytic
transit model of \citet{MAN02}.  The model assumes negligible
eccentricity.  The observations are the R-band data from the 1.8m
Perkins Telescope shown in Figure~\ref{janeslc}.

The calculation has seven free parameters: $M_{\star}$, $R_{\star}$,
$\rho$, $\tau$, $t_{o}$, $u_{1}$, and $u_{2}$.  $t_{o}$ is the
mid-transit time offset from the ephemeris with a period given by the
XO observations and a mid-transit zeropoint near the mid-transit gap
of the light curve, HJD 2454147.75.  The limb darkening coefficients,
$u_{1}$ and $u_{2}$, model the limb darkening with the quadratic law,
$I=1-u_{1}(1-\mu)-u_{2}(1-\mu)^{2}$, where $I$ is the specific
intensity normalized to unity at the center of the stellar disk and
$\mu$ is the cosine of the angle between the line of sight and the
surface normal.  In practice, we follow \citet{HOL06} by adopting
$a_{1}=u_{1}+2u_{2}$ and $a_{2}=2u_{1}-u_{2}$ as the parameters used
in the calculation.  This linear combination of limb darkening
coefficients reduces their mutual degeneracy.

The prior for $R_{\star}$, $\rho$, and $\tau$ is given by
Equation~\ref{eq:unipri}, which is equivalent to a prior uniform in
$R_{\star}$, $R_{p}$, $i$.  The priors for $R_{\star}$, $\rho$, and
$t_{o}$ have cutoff values well beyond values allowed by the data.
The prior on $\tau$ has an upper limit cutoff, $\tau<\tau_{max}$,
where $\tau_{max}$ is the longest transit duration possible for a
given $R_{\star}$, $M_{\star}$, and $\rho$.  We assume uniform priors for the
limb darkening coefficients with the following physically motivated
limits on the parameters.  We require the highest surface brightness
to be located at the disk center ($u_{1}\geq 0.0$), require the
specific intensity to remain above zero ($u_{1}+u_{2}\leq 1.0$), and
do not allow limb-brightened profiles ($u_{1}+2u_{2}\geq 0.0$).  The
form of the prior for $M_{\star}$ is a Gaussian where the central
value and standard deviation of the Gaussian are a function of
$R_{\star}$ as given by the SME isochrone analysis data given in
Table~\ref{table:smeiso} and described in \S~\ref{sec:sme}.  We employ
a spline interpolation over the grid of $M_{\star}$ and $\sigma_{M}$
as a function of $R_{\star}$.

\begin{figure}
\plotone{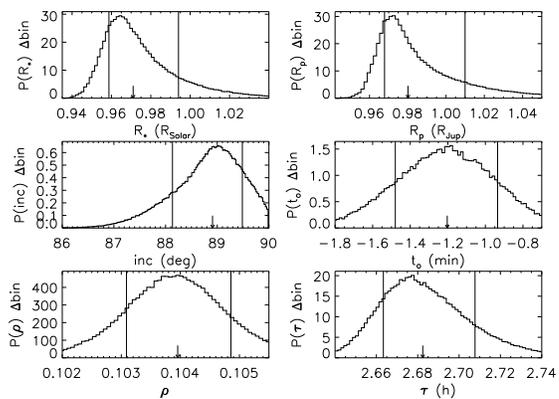}
\caption{Marginalized posterior probability for the \xon\ and \xonb\ parameters from MCMC samples.  We adopt the median of the posterior probability ({\it arrow}) as the point estimate of the parameter.  The {\it solid} lines indicate the 68.3\% credible interval.\label{maspost}}
\end{figure}

The estimate of the posterior probability comes from 7 independent
chains of length $N_{chn}=60000$ with varying initial conditions.
This results in an effective length $N_{eff}=N_{chn}/N_{cor}=12000$
after taking into account the autocorrelation length $N_{cor}=5$.
Using these 7 chains, the largest Gelman-Rubin statistic amongst the
parameters, R=1.0002, where R$<$1.02 indicates convergence of the
chain \citep{GEL92}.  Figure~\ref{maspost} shows the resulting
posterior probability distribution for each parameter after
marginalization over the other parameters.  The posterior probability
is simply a normalized histogram of the MCMC sample values.  We adopt
the median as the best single point estimate of the posterior
probability.  To derive an $\alpha$\% credible interval for a
parameter, the $N$ MCMC samples are sorted by the parameter of
interest.  The lower limit of the credible interval is taken as the
$((1-\alpha)/2)N^{\rm th}$ sorted sample, and the upper limit of the
credible interval is taken as the $(1-(1-\alpha)/2)N^{\rm th}$ sorted
sample.  The arrow point along the abscissa in Figure~\ref{maspost}
indicates the median of the posterior probability, and the the
vertical solid lines in Figure~\ref{maspost} show the 68.3\% credible
interval.

\begin{figure}
\plotone{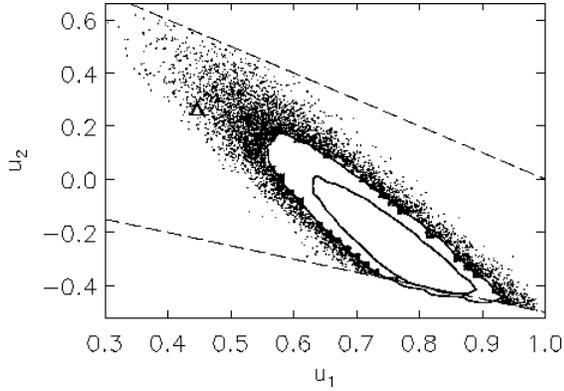}
\caption{ Joint posterior
probability for the stellar limb darkening coefficients, $u_{1}$ and
$u_{2}$.  The
solid contours are isoprobability contours containing 68\% and 90\% of
the MCMC samples.  The remaining 10\% of the samples lying outside the
region of highest probability are also shown ({\it points}).
The region of highest probability differs from the theoretically calculated limb darkening
coefficients ({\it triangle}).  The prior limits for the limb
darkening coefficients are indicated with {\it dashed lines}.\label{limbcontour}}
\end{figure}

Figure~\ref{limbcontour} shows the joint posterior probability for the
stellar limb darkening coefficients, $u_{1}$ and $u_{2}$.  The solid
contours are isoprobability contours containing 68\% and 90\% of the
MCMC samples.  The small points illustrate the remaining 10\% of the
samples lying outside the region of highest probability.  The region
of highest probability for the limb darkening coefficients differs
from the theoretically calculated limb darkening coefficients (open
triangle symbol) obtained from \citet{CLA00}.  The dash lines
illustrate the prior limits for the limb darkening coefficients.  The
lower dash line in Figure~\ref{limbcontour} corresponds to not
allowing limb brightened specific intensity profiles.

\begin{figure}
\plotone{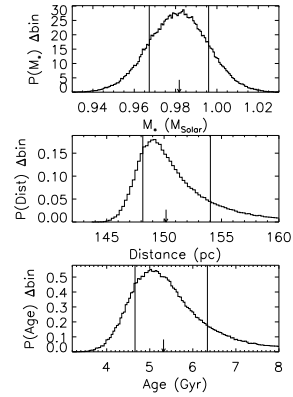}
\caption{The marginalized posterior probability for the \xon\ that rely on the SME isochrone analysis. The SME isochrone analysis translates the posterior distribution for $R_{\star}$ into the posterior distribution for Distance and Age.\label{othpost}}
\end{figure}

Using the SME isochrone analysis, we translate the MCMC samples for
$R_{\star}$ into distance to \xon\ and age estimates.  Similar to the
procedure that defines the prior on $M_{\star}$, we interpolate the
$R_{\star}$ versus trial distance and age relationships
given by the SME isochrone analysis (i.e., for a given stellar radius
estimate, the SME isochrone analysis has a best distance and age
estimate for the system).  The posterior probability for these two
parameters is shown in Figure~\ref{othpost}.  We repeated the
calculation assuming a prior uniform in $\cos{i}$, and this did not
materially affect the parameter estimates.    We
summarize the properties of \xonb\ in
Table~\ref{table:planet}.

The uncertainties given in Table~\ref{table:planet} for \xonb\ and
Table~\ref{table:star} for \xon\ represent the precision of our
experiment.  These uncertainties represent the expected scatter of
values obtained if the experiment was repeated with similar quality
data and identical procedures.  Other systematic sources of error
affect the accuracy of our measurement that are most likely comparable
or larger than our precision.  The sources of systematic error only
enter into our prior for $M_{\star}$.  Our adopted uncertainty in
$M_{\star}\sim$2\% follows from the uncertainty in $T_{eff}$ and
metallicity in the SME isochrone analysis (see \S~\ref{sec:sme}) aided
by the very weak dependence of $M_{\star}$ on the unknown distance to
\xon.  \citet{HIL04} show the Y$^{2}$ isochrones employed in this
study successfully predict stellar masses to within 1\%-3\% for main
sequence stars $M_{\star}>0.6$ \Msun\ with independent dynamical mass
estimates.  For the derived $T_{eff}$ and [Fe/H] for \xon, the
isochrones from \citet{GIR02} at maximum differ by 3\% in $M_{\star}$
from the Y$^{2}$ isochrone prediction.  \citet{COD02} find 7\%
systematic error in $M_{\star}$ for HD 209458 due to uncertainty in
Helium abundance and the treatment of convection.  In light of these
potential sources of systematic uncertainty, we increased the standard
deviation of the Gaussian prior on $M_{\star}$ to $\sigma=0.07$ \Msun.
This larger uncertainty on $M_{\star}$ resulted in more symmetric and
slightly broader posterior distributions for $R_{\star}$ ($\sigma \pm
0.03$ \Rsun), $R_{p}$ ($\sigma \pm 0.03$ \Rjup), and age ($\sigma \pm
1.4$ Gyr).  The surface gravity of the planet, $g_{\rm p}$, is independent
of $M_{\star}$ \citep{SOU07}.  From the measured radial velocity
semi-amplitude and light curve parameters, $g_{\rm p}=$\vgp $\pm$\egp\ \mpstwo\
for \xonb, where the uncertainty in $g_{\rm p}$ is dominated by the uncertainty
in the radial velocity semi-amplitude.  Other parameters weakly
dependent on $M_{\star}$ and available from the light curve
observations directly are $a/R_{\star}=8.2\pm^{0.1}_{0.2}$ and
$a/R_{\rm p}=79.0\pm^{0.9}_{2.5}$.

\subsection{Radial Velocity Measurements}\label{sec:rv}

We measured the mass of \xonb\ using the radial velocity
technique described below and by \citet{MCC06}.
The 2-dimensional spectra obtained with
the HET were extracted to 1-dimensional spectra and associated
approximate wavelength solutions derived from ThAr spectra obtained
during twilight (\S~\ref{sec:spectroscopy}).
To account for optical distortions, the wavelength solution fits the
centroids of thousands of ThAr lines to a function of
the X and Y coordinates of the CCD that includes terms linear in X and
in Y plus the following
cross terms: XY, XXY, XYY, XXYY, XXXY, and XYYY.
Using a downhill simplex $\chi^2$ minimization algorithm, ``Amoeba,'' we
adjusted
the parameters of a synthetic spectrum to fit, in separate
$\sim$15 \AA\ sections, the stellar spectrum observed through an iodine absorption
cell. The requirement of strong iodine lines
limits radial velocity estimates to the wavelength range 5210 $< \lambda
<$ 5700 \AA.
The synthetic spectrum consists of a high-resolution spectrum of the
Sun, the Earth's atmosphere\footnote{Within the
wavelength range of interest, telluric absorption lines are negligible
but are included anyway.}
\citep{WAL98}, and a high-resolution spectrum of an iodine gas
cell \citep{COC00} convolved with
a Voigt profile to approximate the line-spread-function of the
instrument. In addition to the convolution,
we used a few additional free parameters to model specific physical or
instrumental characteristics:
the radial velocity of the star, a shift of the iodine spectrum
attributable to instrumental errors
in the approximate wavelength solution specific to the particular 15 \AA\
section, the continuum
level, and an exponent that adjusts the optical depths of the solar
absorption lines in order to better approximate the
(non-solar) stellar spectrum.
The difference of the parameters representing the radial velocity of the
star and
that attributable to the instrument from the iodine spectrum equals the
topocentric radial velocity of the star, which
we transform to the barycentric frame of the solar system.
We average the individual stellar radial velocity estimates from each of
the 15 \AA\ sections to determine
the stellar radial velocity measurement at each epoch and its associated
1-$\sigma$ uncertainty (see Table~\ref{table:rv}).

\begin{figure}
\plotone{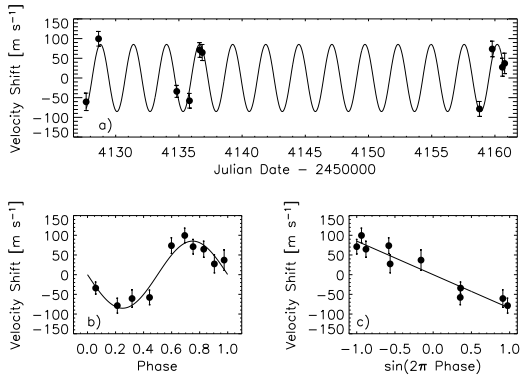}
\caption{a) The radial velocity of \xon\ oscillates sinusoidally with a
semi-amplitude K = \vrvK$\pm$\ervK\ \mps, implying \xonb's mass is
\vMp$\pm$\eMp\ \Mjup. b) The
period and phase of the radial velocities were fixed at values
determined by the transits. The mean stellar radial velocity with
respect to the solar system's barycenter has been subtracted.  In
order to determine K, we used the HET spectra calibrated with an
iodine absorption cell (filled circles).   c) In this representation
of the data, a circular orbit yields a straight line of slope $-$K.
\label{rvnorth}}
\end{figure}

Figure~\ref{rvnorth} shows the resulting radial velocity curve phased
with the \xonb\ ephemeris determined from the transits and assuming
zero eccentricity.  The typical uncertainty for each measurement,
$\sigma_{RV}=20$ \mps.  The radial velocity semi-amplitude,
K=\vrvK$\pm$\ervK\ \mps.  This amplitude results in
$M_{p}=$\vMp$\pm$\eMp\ \Mjup\ for \xonb, assuming $M_{\star}$=\vMs\
\Msun\ for \xon\ and a circular orbit for \xonb.  The six radial
velocities measured for \xon S (Table~\ref{table:rv}) are consistent
within the observational errors with no radial velocity variation and
show no significant evidence for a HJ orbiting \xon S.

\subsection{Ephemeris and Transit Timing Variations}\label{sec:ttv}

The database of E.T.\ transit light curves for \xonb\ enable us to
refine the ephemeris for \xonb\ and to quantify transit timing
variations from this ephemeris.  The transit light from the 1.8m
Perkins Telescope on February 16, 2007 provides a precise measurement
of the mid-transit zeropoint of \xonb.  From the best fitting transit
model in a $\chi^{2}$ sense during the MCMC calculation,
we find the ephemeris zeropoint is
\vjd\ with an uncertainty of 17 s.

To refine the orbital period of \xonb, we calculate mid-transit times
for the nearly complete transit events observed with the XO cameras on
2453355 and 2453376 HJD and the E.T.\ transit events shown in
Figure~\ref{et436} with nearly complete ingress and egress coverage.
We compare the transit data with a limb-darkened transit model.  The
only free parameter of the transit model is the mid-transit time.  The
other parameters of the transit model are fixed at the best fitting
transit model in a $\chi^{2}$ sense during the MCMC calculation, and
the model employs theoretically calculated limb darkening coefficients
obtained from \citet{CLA00} in the appropriate passband of the
observations.  The transit model has a flux decrement of $\rho^{2}/2$
when applied to the XO survey data to account for the transit depth
dilution caused by the flux of \xon S in the photometric aperture.

Table~\ref{table:midpoints} provides the resulting mid-transit times.
The three independent mid-transit estimates on 2454168 HJD enables
estimating the typical uncertainty (1-$\sigma$=3 min) in the transit
timing.  To refine the orbital period we minimize the $\chi^{2}$
difference between the transit timing observations and ephemeris
model.  We keep the ephemeris zeropoint fixed at the value derived
from the 1.8m Perkins Telescope light curve.  The best-fit period,
\vperiod$\pm$\eperiod\ day, results in a $\chi^{2}_{min}=10.1$ for
$\nu$=11 degrees of freedom.  This suggests there are no significant
timing variations.  For the minimization we assumed a uniform error,
$\sigma$=3 min, on the transit timings.  The ephemeris for
\xonb\ accumulates a 5 min uncertainty by 2010.

\section{Discussion}\label{sec:disc}

The transit candidate of \citet{MAN05} illustrates the non-negligible
potential for triple stars to have transit light curves and radial
velocity variations consistent with a planet.  However, in the case of
\xonb, our attempts to explain the light curve and spectroscopy with a
physical stellar triple fail.  We employ the Y$^{2}$ isochrone
appropriate for the the physical properties of \xon\ supplemented with
the low-mass stellar isochrone between 0.072$<M_{\star}<$0.5 \Msun\
from \citet{CHAB00}, stellar limb darkening coefficients from
\citet{CLA00}, and the light curve synthesis routine of \citet{WIL93}
to model a stellar triple system.  The constraints on the transit
duration and transit depth from the light curve require
$M_{\star}>0.95$ \Msun\ for the primary in a stellar binary blended
with the light of \xon.  The required stellar binary has $>$75\% the
flux of \xon\ and has a radial velocity semi-amplitude, $K> 16$ \kps.
Such a binary would be readily apparent in the spectrum of \xon\ given
the narrow spectral features for \xon, $v\sin{i}<2.3$ \kps.  We cannot
completely rule out the possibility of a line-of-sight faint
background binary blended with the light of
\xon\ as an explanation for the observations.  However, the sinusoidal
shape of the radial velocity variations necessitate the line-of-sight
binary to have a systemic velocity similar to \xon\, otherwise the
radial velocity curve develops asymmetries that are not observed
\citep{TOR05}.  Given the large proper motion of \xon, ground-based
adaptive optics or space-based observations have the potential to
definitively rule out a background line-of-sight binary.

In addition to \xon, only HAT-P-1, HD 20782, HD 80606, HD 99492 (GJ
429B), HD 178911B, and HD 186427 (16 Cyg B) have $\Delta V<$2.0 mag
difference between the stellar binary components \citep[][supplemented
by data from the SIMBAD database]{DES07} and can be considered
nearly-equal-mass stellar binaries hosting a known extrasolar planet.
In 5 out of the 7 binary systems, the planet orbits the lower mass
star of the binary.  In the remaining 2 systems, HD 20782 and HD
80606, the planet orbits the more massive star, but the orbits for
these planets have the highest eccentricities amongst all known
planets, e=0.93 and e=0.92, respectively \citep{BUT06}.  Equivalently,
if we restrict the sample of extrasolar planets in nearly-equal-mass
stellar binaries to modest eccentricity (e$<$0.8), in 5 of the 5 known
systems, the planet orbits the lower mass star.  We offer several
speculative possibilities to explain the current pattern of planets in
nearly-equal-mass stellar binaries: 1) The pattern arises due to small
sample statistics, 2) Planets form more readily around the lower mass
component of a nearly-equal-mass stellar binary, 3) Planets form
equally likely around the components of a nearly-equal-mass stellar
binary, but planets orbiting the primary component have lower
detectability due to lower planet mass or higher orbital eccentricity
\citep{CUM04}.

The \xon\ stellar binary has a higher space velocity with respect to
the Local Standard of Rest (LSR), $v=\sqrt{U^2+V^2+W^2}\sim 100$
kms$^{-1}$, than most of the other known extrasolar planets
\citep{SAN03,ECU07}.  According to the purely kinematic classification
of \citet{BEN03} and \citet{BEN06}, \xon\ has $\sim$7 times higher
probability of belonging to the Thick Disk than the Thin Disk.  From a
chemical abundance perspective, a metallicity of [Fe/H]$<-0.3$ is
typically the upper limit for Thick Disk metallicities
\citep{MIS04,RED06}.  However, \citet{BEN06} has recently claimed
kinematically selected Thick Disk stars with [Fe/H]$\sim 0.0$ have a
chemical abundance pattern distinguishable from Thin Disk stars at the
same metallicity.  The sample of stars with kinematics suggestive of
the Thick Disk but Thin Disk metallicities, like \xon, is still too
small to determine their relationship to the traditional Thin/Thick
Disk populations.  However, \citet{MIS04} distinguishes between objects with
Thick Disk kinematics and Thin Disk metallicity by determining the
maximum height above the Galactic disk their orbit attains, $Z_{max}$.
Both the estimate of \citet{BEN05} for $Z_{max}$ and integrating the
orbit of \xon\ using the axi-symmetric, static Galaxy potential model
of \citet{ALL91} yield $Z_{max}\sim 100$ pc for the orbit of \xon.  Given
the much larger scale height for the Thick Disk, $\sigma_{TD}\sim 1000$
kpc, under the classification of \citet{MIS04}, \xon\ belongs to the
Thin Disk, since its orbit is well confined to the Galactic plane.
The high space velocity of \xon\ results from high eccentricity rather
than excursions from Galactic plane.  Using the simple model from
\citet{ALL91}, the orbit of \xon\ has a pericenter within $\sim$4 kpc of the
Galactic center and apocenter $\sim$9 kpc (e$\sim$0.4).

Currently, \xon\ lags the LSR by $V=-78$ kms$^{-1}$ and is moving away
from the Galactic center at $U=-71$ kms$^{-1}$.  This places \xon\ in
phase space near the Hercules stream
\citep{FAM05,ECU07} ($U=42$, $\sigma_{U}=28$, $V=-52$, $\sigma_{V}=9$
kms$^{-1}$), but is lagging the LSR more than the typical Hercules
stream member.  The possibility that \xon\ belongs to a dynamical
stream originating from the metal-rich inner Galaxy is an
example that supports the findings of \citet{FAM05} and \citet{ECU07}
that compared to field stars, metal-enhanced dynamical streams should
be over-abundant in detectable extrasolar planets.

\section{Summary}\label{sec:sum}

The star \xon, GSC 34130-0005, hosts an approximately Jupiter-size, 0.6 Jupiter-mass transiting extrasolar planet, \xonb, with an orbital
period $\sim$2.6 days.  \xon\ is a V=11.2 early K dwarf with high
metallicity, [Fe/H]=\vFeH, high proper motion, $\mu_{tot}=157$
mas~yr$^{-1}$, and has a common proper motion stellar companion with
\vSepas $\arcsec$ separation.  The followup high resolution
spectroscopy yields $M_{\star}$=\vMs $\pm$\eMs\ \Msun\ for the mass
of \xon\ and $M_{p}$=\vMp$\pm$\eMp\ \Mjup\ for the mass of \xonb.
The followup high precision photometry yields
$R_{\star}$=\vRs$\pm^{\epRs}_{\enRs}$ \Rsun\ for the radius of \xon\
and $R_{p}$=\vRp$\pm^{\epRp}_{\enRp}$ \Rjup\ for the radius of
\xonb.  Joint analysis of the light curve and spectroscopy yields an
isochrone based age of \xon\, $t$=\vAge$\pm^{\epAge}_{\enAge}$ Gyr,
and an isochrone based distance to \xon\,
$d$=\vDs$\pm^{\epDs}_{\enDs}$ pc.  The quoted values and their uncertainties are
Bayesian credible intervals encompassing 68\% of the marginalized
posterior probability.  For the Bayesian analysis we assume a
Gaussian-like prior on $M_{\star}$, derived from the spectroscopic and
isochrone analysis of
\xon\, and uniform priors on $R_{\star}$, $R_{p}$, orbital
inclination, and limb darkening coefficients.  We describe an
efficient Markov Chain Monte Carlo method to calculate the Bayesian
posterior probability for the system parameters from a transit light
curve.

\xonb\ adds to the sample of Jupiter-mass planets residing in
nearly-equal-mass stellar binaries.  In 5 of the 7 nearly-equal-mass
stellar binaries hosting a planet, the planet orbits the lower mass
stellar component.  In the remaining 2 stellar binaries where the
planet orbits the higher mass star, the planets' orbits are highly
eccentric (e$>$0.9).  Equivalently, if we restrict the sample of
extrasolar planets with modest eccentricity (e$<$0.8) orbits in
nearly-equal-mass stellar binaries, in 5 of the 5 known systems, the
planet orbits the lower mass star.  We speculate on possible
astrophysical reasons for this pattern beyond the simple fact that the
statistics are based upon a very small sample and thus may be
spurious.

With its high proper motion, \xon\ has kinematics suggestive of Thick
Disk membership but a Thin Disk metallicity.  In contrast to the typical
Thick Disk member, the high proper motion of \xon\ results from an
eccentric orbit (Galactic pericenter, $R_{per}<4$ kpc) well confined
to the Galactic disk, $Z_{max}\sim 100$ pc.  \xon\ may originate in
the metal-rich inner Thin Disk and was dynamically scattered into the
solar neighborhood.  Similar to the findings of
\citet{FAM05} and \citet{ECU07}, the discovery of
\xonb\ suggests metal-enhanced dynamical streams from the inner Galaxy
may be abundant in detectable extrasolar planets.

\acknowledgments

The University of Hawaii staff have made the operation on Maui possible; we
thank especially Jake Kamibayashi, Bill Giebink, Les Hieda,
Jeff Kuhn, Haosheng Lin, Mike Maberry, Daniel O'Gara, 
Joey Perreira, Kaila Rhoden, and the director of the IFA, Rolf-Peter Kudritzki.
The Hobby-Eberly Telescope (HET) is a joint project of the University of Texas at Austin, the Pennsylvania State University, Stanford University, Ludwig-Maximilians-Universit\"{a}t M\"{u}nchen, and Georg-August-Universit\"{a}t G\"{o}ttingen.  The HET is named in honor of its principal benefactors, William P. Hobby and Robert E. Eberly.  We thank the HET night-time and day-time support staff and the Resident Astronomer telescope operator; we especially thank John Caldwell, Frank Deglman, Heinz Edelmann, Stephen Odewahn, Vicki Riley, Sergey Rostopchin, Matthew Shetrone, and Chevo Terrazas.

We thank Dave Healy, Lisa Prato, and Marcos Huerta for assistance
observing.  We acknowledge helpful discussions with Julio Chaname, Ron
Gilliland, and Zheng Zheng.  We thank the referee for the insightful
suggestions and significant improvements to the manuscript.

This research has made use of the SIMBAD database, operated at CDS, Strasbourg, France;
data products from the Two Micron All Sky Survey (2MASS),
the Digitized Sky Survey (DSS), and The Amateur Sky Survey (TASS);
source code for transit light-curves (Mandel \& Agol 2002);
and community access to the HET.

XO is funded primarily by NASA Origins grant NNG06GG92G and the Director's
Discretionary Fund of the STScI.

\appendix
\section{Markov Chain Monte Carlo Details}\label{apx:mcmc}
Care must be taken to properly assign prior probabilities for a
nonlinear transformation in parameters \citep{CHU03}.  In particular,
a uniform prior in $i$ is not the same as a uniform prior in $\tau$.
The transformation law of probabilities provides the necessary prior
probability to maintain the uniform prior in $R_{\star}$, $R_{p}$, and
$i$ for our chosen set of free parameters: $R_{\star}$, $\rho$, and $\tau$, where

\begin{equation}
\tau=\frac{P}{\pi}{\rm arcsin}\left(\frac{R_{\star}}{a}\left[\frac{(1+\rho)^{2}-(a/R_{\star}\cos{i})^{2}}{1-\cos{i}^{2}}\right]^{1/2}\right),
\label{eq:tau}
\end{equation}
where $P$ is the orbital period, $a$ is the semi-major axis, and we
have assumed zero eccentricity \citep{SEA03}.

The general transformation law
with multiple dimensions states
\begin{equation}
p(y_{1},y_{2},...)dy_{1}dy_{2}...=p(x_{1},x_{2},...)\left\| \frac{\partial (x_{1},x_{2},...)}{\partial (y_{1},y_{2},...)} \right\| dy_{1}dy_{2}...,
\end{equation}
where the original joint probability distribution,
$p(x_{1},x_{2},...)dx_{1}dx_{2}...$, is transformed into another
probability distribution in terms of the new set of variables
$(y_{1},y_{2},...)$ by multiplication with the absolute value of the
Jacobian determinant, $\| \partial ()/\partial () \| $.  The new $y$
variables must have the same number and be expressible in terms of the
old $x$ variables \citep{PRE92}.  Writing the old variables in terms of the new variables, $R_{\star}=R_{\star}^{\prime}$, $R_{p}=\rho R_{\star}^{\prime}$, and
\begin{equation}
i=\arccos{\sqrt{\frac{(R_{\star}^{\prime}/a)^{2}(1+\rho)^{2}-\sin{\phi}^2}{(1-\sin{\phi}^{2})}}},
\end{equation}
where $\phi=\tau \pi/P$, the Jacobian matrix is written
\begin{equation}
\| \partial ()/\partial () \|=\left\| \begin{array}{ccc}
\frac{\partial R_{\star}}{\partial R_{\star}^{\prime}}=1 & \frac{\partial R_{\star}}{\partial \rho}=0 & \frac{\partial R_{\star}}{\partial \tau}=0 \\
\frac{\partial R_{p}}{\partial R_{\star}^{\prime}} & \frac{\partial R_{p}}{\partial \rho} & \frac{\partial R_{p}}{\partial \tau}=0 \\
\frac{\partial i}{\partial R_{\star}^{\prime}} & \frac{\partial i}{\partial \rho} & \frac{\partial i}{\partial \tau} \\
\end{array}\right\|,
\end{equation}
where we have indicated the trivial elements with value one or zero.  Overall, the Jacobian simplifies to
\begin{equation}
| (\partial R_{p}/\partial \rho)(\partial i/\partial \tau) |=\frac{\pi}{P} R_{\star}^{\prime}\sin{\phi}\sqrt{\frac{((R_{\star}^{\prime}/a)^{2}(1+\rho)^{2}-1)}{(\sin{\phi}^{2}-(R_{\star}^{\prime}/a)^{2}(1+\rho)^{2})}}.
\label{eq:unipri}
\end{equation}
With our choice of parameters, the proper prior to maintain uniform
priors in $R_{\star}$, $R_{p}$, and $i$ is the above Jacobian
multiplied by the original joint probability distribution for uniform
priors, which is a constant.  The prior probability goes to zero for $0<
\tau <
\tau_{max}$, where $\tau_{max}=P/\pi \arcsin{(R_{\star}/a)(1+\rho)}$.

A trial chain is necessary to define the covariance matrix for the
final eigenbasis set of parameters.  $M_{\star}$, $R_{\star}$, $\rho$,
$\tau$, $u_{1}$, and $u_{2}$ enter into the covariance matrix.  We
find $t_{o}$ shows no significant correlation with respect to the
other parameters.  To determine the necessary linear transformation,
we determine the eigenvectors of the covariance matrix built from a
trial chain \citep{TEG04}.  This is equivalent to a Principal
Component Analysis.  The bottom panels in Figure~\ref{tranfig} show
trial samples in terms of the eigenvectors of the covariance matrix.
The eigenvector along the largest variance, $e_{1}$, is predominately
along $R_{\star}$, and $e_{2}$ and $e_{3}$ are a mixture of $\rho$ and
$\tau$.

The autocorrelation length of samples in the MCMC sequence is one
method to quantify the efficiency of the calculation.  With the
original set of parameters ($R_{\star}$, $R_{p}$, and $i$), the
correlation length (when the autocorrelation drops by half) in the
$R_{\star}$ parameter varied, $500<N_{cor}<1500$ steps.  Using our new
set of parameters ($R_{\star}$, $\rho$, and $\tau$), $40<N_{cor}<60$
steps.  Finally, with the eigenbasis parameters, $4<N_{cor}<5$ steps,
where the correlation is measured in the physical variable
$R_{\star}$.

\clearpage

\clearpage

\begin{deluxetable}{cccccc}
\tabletypesize{\small}
\tablewidth{0pt}
\tablecaption{{\rm XO Survey \& E.T.\ Light Curve Data}\tablenotemark{a}}
\startdata
\hline
\hline
Heliocentric Julian Date & Light Curve & Uncertainty & Filter & N\tablenotemark{b} & Observatory \\
                         &  [mag]      & (1-$\sigma$) [mag] & & & \\
\hline
2452961.14380 & -0.0027 & 0.0039  &   W &  1 &  XO \\
2452961.14404 & -0.0002 & 0.0037  &   W &  1 &  XO \\
2452964.11621 & -0.0057 & 0.0034  &   W &  1 &  XO \\
2452964.11646 & -0.0018 & 0.0033  &   W &  1 &  XO \\
2452964.12329 &  0.0034 & 0.0034  &   W &  1 &  XO \\
\hline
\enddata
\tablenotetext{a}{The complete version of this table is in the electronic edition of
the Journal.  The printed edition contains only a sample.}
\tablenotetext{b}{Average of N measurements}
\label{table:lc}
\end{deluxetable}

\begin{deluxetable}{lccl}
\tabletypesize{\small}
\tablewidth{0pt}
\tablecaption{{\rm Stellar Properties}}
\startdata
\hline
\hline
Parameter & \xon N & \xon S & Reference\\
\hline
GSC ID        & 03413-00005                    & 03413-00210                     & a\\
RA (J2000.0)  & $ 7^h48^m06^s.47 $             & $ 7^h48^m07^s.48 $              & a,b \\
Dec (J2000.0) & +50\arcdeg13\arcmin33\arcsec.0 &  +50\arcdeg13\arcmin03\arcsec.3 & a,b \\
Galactic Latitude b [deg]       & 29.33        &  ...                            & a \\
  ''     Longitude l [deg]      & 168.29       &  ...                            & a \\
V             & 11.18$\pm$0.03                 &  11.12$\pm$0.03                 & c \\
(B-V)         & 0.82$\pm$0.05                  &  0.79$\pm$0.05                  & c \\
(V-R$_{c}$)     & 0.49$\pm$0.05                  &  0.46$\pm$0.05                  & c \\
(V-I$_{c}$)     & 0.86$\pm$0.05                  &  0.82$\pm$0.05                  & c \\
V$_{T}$             & 11.24                          & 11.20                           & b,d \\
(B-V)$_{T}$         & 0.70                           & 0.86                            & b,d \\
J             & 9.74$\pm$0.02                  & 9.74$\pm$0.02                  & e \\
(J-H)         & 0.40$\pm$0.03                  & 0.37$\pm$0.03                   & e \\
(H-K)         & 0.03$\pm$0.03                  & 0.10$\pm$0.03                  & e \\
Spectral Type & \sptype                        &  \sptype                        & c \\
Distance [pc] & \vDs$\pm^{\epDs}_{\enDs}$               &  ...                            & c \\
$\mu_{\alpha}$ [mas\ yr$^{-1}$] & -34.7$\pm$2.6& -33.1$\pm$2.9                   & b \\
$\mu_{\delta}$ [mas\ yr$^{-1}$] &-153.6$\pm$2.4& -154.1$\pm$2.7                  & b \\ 
Total $\mu$[mas\ yr$^{-1}$]   & 157            & 158                              & f \\
Radial Velocity (Bary) [km\ s$^{-1}$] & 47.4$\pm$0.5   & ...                         & c \\
U Space Velocity [km\ s$^{-1}$] & -72.0        &  ...                            & b,c,g,h \\
V ''                           & -78.0        &  ...                            &  b,c,g \\
W ''                            & -4.6         &  ...                            & b,c,g \\ 
Stellar Mass [$M_{\odot}$] &  \vMs$\pm$\eMs       & ...                              & c \\
Stellar Radius [$R_{\odot}$] & \vRs$\pm^{\epRs}_{\enRs}$    & ...                       & c \\
\enddata
\\
References:\\
a) SIMBAD\\
b) Tycho-2 \citep{HOG00} \\
c) this work \\
d) Tycho-2 photometry on the Johnson system \citep{BES00,MAM02}\\
e) 2MASS \citep{SKR06} \\
f) LSPM \citep{LEP05} \\
g) Space Velocity w.r.t. LSR after correction for Solar peculiar motion \citep{MIH81}. \\
h) Negative U is away from the Galactic Center \\
\label{table:star}
\end{deluxetable}

\begin{deluxetable}{cccc}
\tabletypesize{\small}
\tablewidth{0pt}
\tablecaption{{\rm Radial Velocity Shifts}}
\startdata
\hline
\hline
Object & Julian Date & Radial Velocity &  Uncertainty \\
       &   -245000   &  Shift [\mps] &  (1-$\sigma$) [\mps]  \\
\hline
\xon N &           &           &       \\
     &  4127.6626  &    -60.6  &    22 \\
     &  4128.6459   &    99.5    &  18 \\
     &  4134.8318   &   -34.0    &  15 \\
     &  4135.8366   &   -57.8    &  19 \\
     &  4136.6494   &    71.0    &  19 \\
     &  4136.8477   &    64.6    &  20 \\
     &  4158.7782   &   -78.5    &  19 \\
     &  4159.7882   &    73.6    &  20 \\
     &  4160.5891   &    27.3    &  23 \\
     &  4160.7715   &    37.0    &  26 \\
\xon S &            &           &      \\
       & 4133.6383  &    -21.6  &    18 \\
       & 4134.6257  &    -1.5   &   15 \\ 
       & 4135.8481  &    -32.4  &    20 \\ 
       & 4136.8599  &    -7.6   &   17 \\
       & 4158.7932  &     30.6  &    17 \\
       & 4168.7572  &     16.7  &    17 \\
\enddata
\label{table:rv}
\end{deluxetable}

\begin{deluxetable}{ccccc}
\tabletypesize{\small}
\tablewidth{0pt}
\tablecaption{{\rm Results of the SME Analysis}}
\startdata
\hline
\hline
Parameter & \xon N &  \xon S & Uncertainty (1-$\sigma$) & Uncertainty (99.7\%) \\
\hline
$T_{eff}$ [K]	&5340  &  5500 & 32     & 233 \\
log$g$ [\cmpss]	&4.48  &  4.62 & 0.05	& 0.36 \\
$v$~sin~$i$ [\kps]&1.4   &  1.2  & 0.3	& 2.1 \\
\ [M/H]     	&0.44  &  0.45 & 0.02	& 0.20 \\
\ [Na/H]	&0.49  &  0.63 & 0.02	& 0.18 \\
\ [Si/H]	&0.39  &  0.47 & 0.02	& 0.12 \\
\ [Ti/H]	&0.36  &  0.42 & 0.04	& 0.26 \\
\ [Fe/H]	&0.45  &  0.47 & 0.02	& 0.22 \\
\ [Ni/H]	&0.50  &  0.52 & 0.02	& 0.16 \\
\ [Si/Fe]   	&-0.06 &  0.00 & 0.03	& 0.25 \\
\enddata
\label{table:sme}
\end{deluxetable}

\begin{deluxetable}{lccc}
\tabletypesize{\small}
\tablewidth{0pt}
\tablecaption{{\rm Spectroscopically Derived Stellar parameters}}
\startdata
\hline
\hline
Parameter	&	@ 140 pc&	@ 150 pc&	@ 170 pc\\
\hline
   		&	0.98	&	0.97	&	0.96	\\
Mass [\Msun]   	&	0.99	&	0.98	&	0.98	\\
    		&	1.00	&	1.00	&	0.99	\\
    		&	    	&	    	&	    	\\
    		&	0.90	&	0.95	&	1.07	\\
Radius [\Rsun] 	&	0.91	&	0.97	&	1.10	\\
    		&	0.93	&	0.98	&	1.12	\\
    		&	    	&	    	&	    	\\
    		&	4.49	&	4.42	&	4.32	\\
Log(g) [\cmpss] &	4.51	&	4.45	&	4.34	\\
    		&	4.53	&	4.48	&	4.37	\\
    		&	    	&	    	&	    	\\
    		&	0.77	&	3.55	&	8.35	\\
Age [Gyr]     	&	2.04	&	5.27	&	9.49	\\
    		&	3.68  	&	6.92	&	10.62	\\
\enddata
\\
For each parameter, the middle row is the maximum likelihood value, and the\\
values in the rows above and below span the 68\% likelihood of the probability\\
distributions (cf. Figure~\ref{fig:smeiso}).  The three columns correspond to three assumed\\
distances for \xon.\\
\label{table:smeiso}
\end{deluxetable}

\begin{deluxetable}{cc}
\tabletypesize{\small}
\tablewidth{0pt}
\tablecaption{{\rm Mid-Transit Times}}
\startdata
\hline
\hline
Heliocentric Julian Date\tablenotemark{a} & Observatory ID\tablenotemark{b} \\
 -2450000                &                              \\
\hline
3355.14209 & XO \\
3376.07349 & XO \\
4126.82324 & MF \\
4126.82324 & BG \\
4134.66992 & MF \\
4142.51660 & EM \\
4147.75049 & BG \\
4160.82812 & BG \\
4168.67578 & BG \\
4168.68066 & CF \\
4168.67871 & MF \\
\enddata
\tablenotetext{a}{Based upon independent observations on 2454168 HJD, uncertainty in mid-transit time, $\sigma$=3 min.}
\tablenotetext{b}{Observatory ID is author initials, except XO is XO survey data.}
\label{table:midpoints}
\end{deluxetable}

\begin{deluxetable}{lc}
\tabletypesize{\small}
\tablewidth{0pt}
\tablecaption{{\rm The Planet \xonb}}
\startdata
\hline
\hline
Parameter & Value \\
\hline
$P $ 				& \vperiod$\pm$\eperiod\ d 		\\
$t_c $	 			& \vjd$\pm$\ejd\ (HJD)			\\
$K $ 				& \vrvK$\pm\ervK$\ \mps 	 	 \\
$a $ 				& \vap$\pm$\eap\ A.U. 			 \\
$i $  		                & \vincl$\pm$\eincl\ deg  		 \\
$M_{\rm p} $ 			& \vMp$\pm$\eMp\ \Mjup		 	\\
$R_{\rm p} $ 			& \vRp$\pm^{\epRp}_{\enRp}$ \Rjup       \\
$g_{\rm p} $                    & \vgp$\pm$\egp           \mpstwo        \\
\enddata
\label{table:planet}
\end{deluxetable}

\clearpage

\end{document}